\documentclass[twocolumn,superscriptaddress,showpacs,showkeys,preprintnumbers,aps,prl]{revtex4}

\usepackage{nicefrac}
\usepackage{amssymb}
\usepackage{amsmath}
\usepackage{graphicx}
\usepackage{float}
\usepackage{color}

\begin{document}
\title{\boldmath Temperature-dependent charge transport in the
  compensated ferrimagnet Mn$_{1.5}$V$_{0.5}$FeAl from first principles
}
\author{R.~Stinshoff}
\affiliation{Max-Planck-Institut f\"ur Chemische Physik fester Stoffe,
  01187  Dresden, Germany}
\author{S.~Wimmer}
\author{H.~Ebert}
\affiliation{Ludwig-Maximilians-Universit\"at, Dept. Chemie,
  Butenandtstr.~11, 81377 M\"unchen, Germany}
\author{G.~H.~Fecher}
\author{C.~Felser}
\author{S.~Chadov}
\affiliation{Max-Planck-Institut f\"ur Chemische Physik fester Stoffe,
  01187  Dresden, Germany}
\email{stanislav.chadov@cpfs.mpg.de}
\begin{abstract}
  We present an {\it ab-initio} study of the  temperature-dependent  longitudinal and anomalous Hall resistivities in
  the compensated collinear ferrimagnet Mn$_{1.5}$V$_{0.5}$FeAl. Its
   transport properties are calculated using
  the general fully relativistic Kubo--Bastin formalism and their temperature
  dependency is accounted for  magnetic and structural disorder. Both scattering sources, together with the
  residual chemical disorder, were treated equally provided by
  the CPA (Coherent Potential Approximation) SPR-KKR (Spin-Polarized Relativistic Korringa-Kohn-Rostoker) method. All calculated properties showed good agreement
  with a recent experimental results, providing useful specific information on the chemical
  and magnetic arrangement as well as on the influence of
  disorder. Finally, we demonstrated that the anomalous Hall effect in
  such compensated systems occurs regardless of the vanishing net spin moment.
\end{abstract}
\maketitle
Magnetically compensated systems provide an attractive base for the
next generation of spintronic devices~\cite{MT11}. Their investigation is motivated by
 potential applications in various technological fields, such as new types of RAM, detectors,
microscopic tips, etc, in which the interest is focused on an alternative
manipulation of spins, absence of stray fields and higher
operating frequencies.  Magnetically compensated systems have different
order parameters than ferromagnets, such as staggered
magnetization~\cite{BCM+13,WHZ+16} or magnetic
chirality~\cite{BTD13,WHO16,SSN+16}, which can be manipulated and
detected by either magnetic fields or pulsed electric currents. However,
the absence of net magnetization does not exclude the possibility that
such materials will exhibit the
anomalous Hall effect (AHE)~\cite{KF14}, Kerr effect~\cite{FGZ+15} or high
spin-polarization~\cite{CKF13,WFCF15,CSW+15}. For example, in case of
the planar noncollinear antiferromagnets (e.g., Mn$_3$Ir) AHE has been predicted~\cite{CNM14}
  for the case when the mirror symmetry is broken. 
By considering  magnetic compensation in the cubic ferrimagnets, it is important
to note that, both typical cubic structures with {\it
  Fm\,\,$\bar{\text{\!\!3}}$m} or {\it  F~$\bar{\text{\!\!4}}$3m} space
groups correspond to {\it I4/mm$'$m$'$} or {\it
  I\,\,$\bar{\!\!\text{4}}$m$'$\!2$\,'$} magnetic space groups,
respectively. Both cases belong to the magnetic Laue group {\it
  4/mm$'$m$'$}~\cite{Kle66,SKWE15} which leads to  the following shape of the conductivity
tensor:
\begin{eqnarray}
  \underline{\sigma} =\left(\begin{array}{rcc}\sigma_{\rm xx} & \sigma_{\rm H} & 0
    \\ -\sigma_{\rm H} & \sigma_{\rm xx} & 0 \\ 0 & 0 & \sigma_{\rm zz} \end{array}\right),
\end{eqnarray}
where $\sigma_{\rm H}$ is the anomalous Hall component.
Obviously, $\sigma_{\rm H}$ will have a non-vanishing amplitude if there
is a difference between the spin-up and -down projections of the
electronic structure, which can be fulfilled if the magnetic
moment of one atom type is compensated by the antiparallel moments from the
other atom types. It is particularly easy to realize such systems using cubic Heusler alloys since most of them obey the
Slater--Pauling rule~\cite{Sla36,Pau38}, suggesting that compensated ferrimagnets
can be found among compounds having 24 electron formula
units. Some compensated Heusler ferrimagnets  have been already
reported, such as MnCo$_{\nicefrac{4}{3}}$Ga$_{\nicefrac{5}{3}}$~\cite{LLZ+12}. Ferrimagnetic compensation can 
 also be induced  in the tetragonal structures~\cite{WCKF15}, 
e.g., in the case of Mn$_{1.4}$Pt$_{0.6}$Ga~\cite{NNC+15}; however, the deviation
from the Slater--Pauling rule  does not allow for a clear recipe for the
exact compensating stoichiometry. 

The first experimental evidence of non-zero AHE
in compensated cubic ferrimagnets was given
recently~\cite{SNF+17,SFC+17} for the 
Heusler compound Mn$_{1.5}$V$_{0.5}$FeAl. Additional calculations~\cite{SFC+17} have
 shown that this system is half-metallic in agreement with the
Slater-Pauling rule, indicating that the observed AHE is due to the aforementioned strong asymmetry of the
spin-channels. Here, we investigate this scenario by first-principles
calculations on the system Mn$_{1.5}$V$_{0.5}$FeAl and verify that the experimental
non-zero AHE is an intrinsic property of the compensated ferrimagnets,
rather than a consequence of the small remaining magnetization 
induced by deviations from stoichiometry. We employed the fully-relativistic 
SPR-KKR (Spin-Polarized Relativistic Korringa-Kohn-Rostoker) method using the
standard generalized gradient approximation~\cite{PBE96} for the
exchange-correlation potential. The structural information on
Mn$_{1.5}$V$_{0.5}$FeAl is taken from a recent experiment~\cite{SNF+17}.

Though the origins of AHE being well
understood theoretically, a realistic combined first-principles
 description still remains a challenging
computational task. At present, the most general approach for equally
considering the sources of AHE is the so-called
Kubo-Bastin formalism. Being implemented within the SPR-KKR
method~\cite{EKM11,KCE15}, it allows us to deal with the
charge transport in solids by treating various disorder effects
on the basis of the CPA (Coherent Potential
Approximation)~\cite{Sov67,Tay67}. 

\begin{figure}
\centering
  \includegraphics[width=0.9\linewidth,clip]{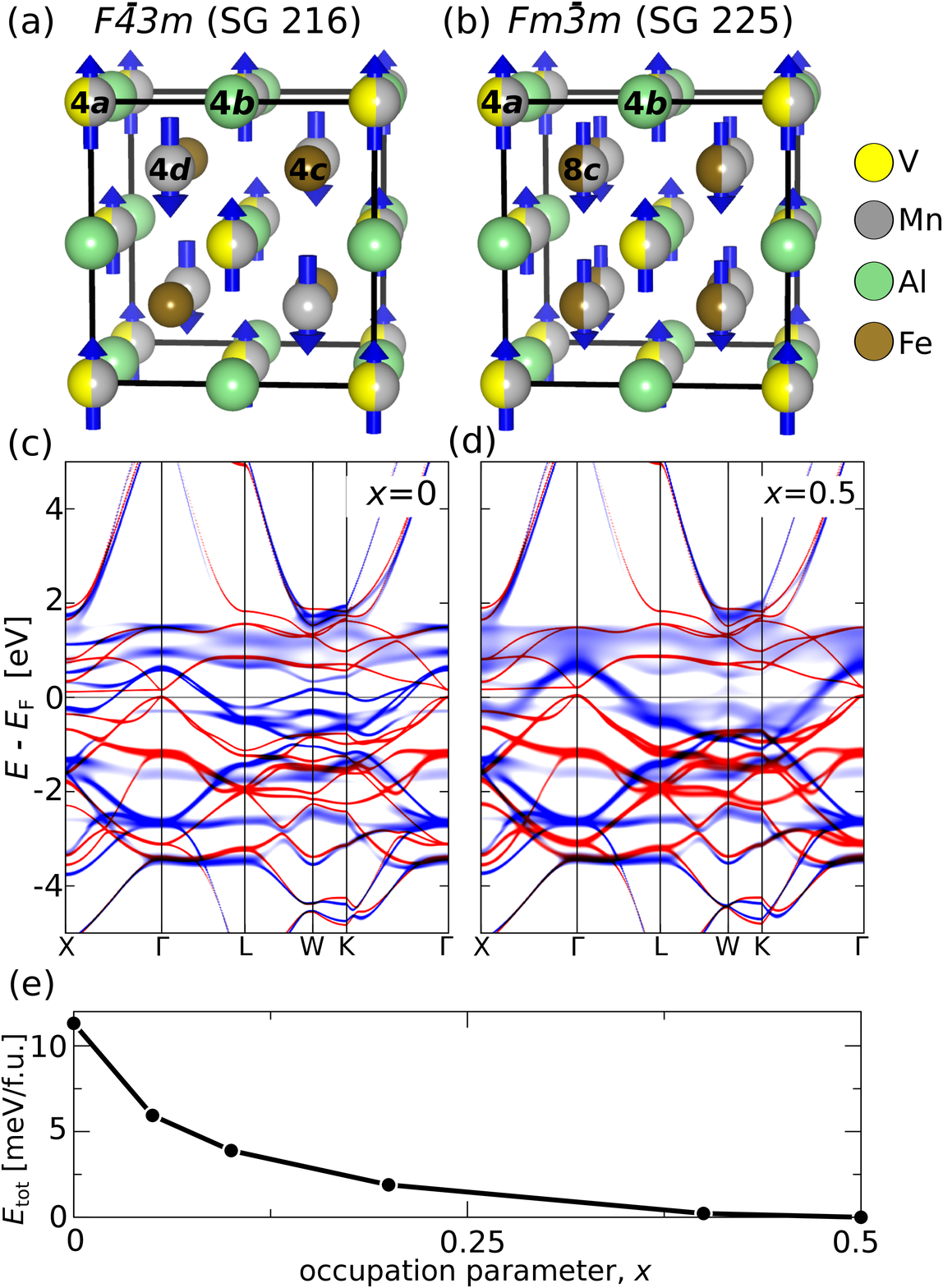}
 \caption{Mn$_{1.5}$V$_{0.5}$FeAl within (a)~{\it F~$\bar{\text{\!\!4}}$3m} (SG\,216,
   with  $4c$ and $4d$ Wyckoff sites occupied by Mn and Fe,
   respectively) and (b)~{\it Fm\,\,$\bar{\text{\!\!3}}$m} (SG\,225, for which $4c$ and
   $4d$ sites  become equivalent by changing to common type $8c$
   with random Mn$_{0.5}$Fe$_{0.5}$ occupation). Other sites, $4a$ and $4b$, occupied by Mn$_{0.5}$V$_{0.5}$ and
   Al, respectively, remain the same in both structures. Arrows indicate
   the spin moments of Mn atoms. (c) and (d)~show the corresponding (to (a) and (b), respectively) spin-resolved (red - spin-up, blue
   - spin-down) spectral densities. (e)~The total energy as a
   function of $x$ (occupation parameter), i.e., the amount of Mn in $4d$ position: ${x=0}$ corresponds
   to (a), ${x=0.5}$ - to (b).} \label{FIG:structures}
\end{figure}

Since the X-ray diffraction (XRD) refinement~\cite{SNF+17} does not
unambiguously resolve the occupancies of
the $4c$ and $4d$ Wyckoff positions, we first specified the chemical order in the system. Most of the integral
characteristics of the system, such as the magnetization, are not very sensitive to
the partial ordering; however partial ordering might significantly influence the charge
transport. Treating our system within the {\it F~$\bar{\text{\!\!4}}$3m}
symmetry, we assumed $4c$ and $4d$ sites were different. This
allowed us to mix Mn with Fe, gradually going from the
most ordered case (Mn)$_{4d}$(Fe)$_{4c}$ ({\it F~$\bar{\text{\!\!4}}$3m}) to the most
disordered case (Mn$_{0.5}$Fe$_{0.5}$)$_{4d}$(Mn$_{0.5}$Fe$_{0.5}$)$_{4c}$, which 
has higher effective symmetry ({\it Fm\,\,$\bar{\text{\!\!3}}$m}). Both variants are shown in Fig.\,\ref{FIG:structures}\,a and b.
Even though their electronic structures
(Fig.\,\ref{FIG:structures}\,c, d) were looking similar, increased
broadening of the spin-down states was observed in the vicinity of the Fermi energy $E_{\rm F}$  for the
case (d), which was caused by the additional Mn/Fe disorder. This
broadening  should impose a
drastic difference in the transport properties of the case (d) with respect to case (c).
Calculating the total energy as a function
of the occupation rate $x$:
(Mn$_{x}$Fe$_{1-x}$)$_{4d}$(Mn$_{1-x}$Fe$_{x}$)$_{4c}$, ${0\le x\le
  0.5}$, provided information concerning the most stable phase. 
As shown in Fig.\,\ref{FIG:structures}\,e, the total energy decreased 
monotonically with $x$ and reached its minimum at ${x=0.5}$. This behavior
indicates that Mn$_{1.5}$V$_{0.5}$FeAl effectively has {\it Fm\,\,$\bar{\text{\!\!3}}$m} symmetry.

\begin{figure}
\centering
  \includegraphics[width=0.9\linewidth,clip]{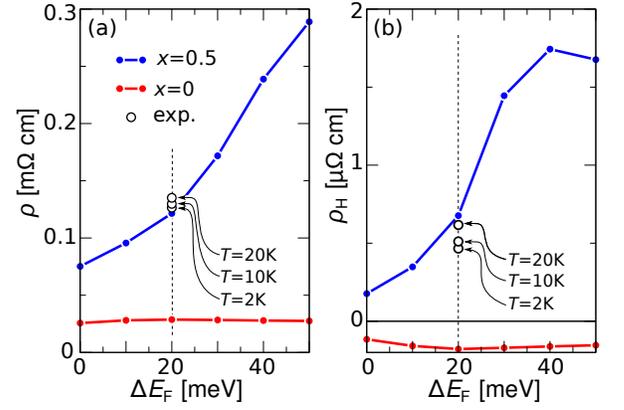}
 \caption{Residual resistivities as a function of the Fermi energy
   shift ${\Delta E_{\rm F}=E'_{\rm F}-E_{\rm F}}$. (a)~Longitudinal
   resistivity $\rho$, (b)~anomalous Hall resistivity $\rho_{\rm
     H}$. Red and blue curves correspond to the ${x=0}$ and ${x=0.5}$
   variants; empty circles are the experimental values~\cite{SFC+17} measured at low temperatures (indicated explicitly).\label{FIG:efcalibration}} 
\end{figure}

Having specified the chemical order, we proceeded with
the precise calibration of the Fermi energy $E_{\rm F}$. Again, small
deviations of $E_{\rm F}$ do not influence the integral properties
as the magnetization, but might be crucial for the transport
properties. These deviations can occur both in 
experiment (e.g., due to chemical and structural imperfections) as well as
in calculations (e.g., due to the spherical approximation of the
atomic potentials). For this reason, we computed both
${\rho=\nicefrac{1}{3}\cdot(2\rho_{\rm xx}+\rho_{\rm zz})}$ and ${\rho_{\rm H}=\rho_{\rm xy}}$ as functions of
the $E_{\rm F}$ position~(Fig.\,\ref{FIG:efcalibration}). 
For ${x=0.5}$ we observed a strong dependence on $E_{\rm F}$ for both ${\rho}$ and ${\rho_{\rm H}}$, which
showed the best simultaneous agreement with experiment at ${\sim20}$\,meV above the nominal $E_{\rm F}$. At the same time, for ${x=0}$ 
 both quantities strongly deviated from experiment within the whole range of ${\Delta E_{\rm F}}$.

The temperature dependency of the
charge transport for ${x=0.5}$ was examined by considering two basic
sources  of disorder induced by temperature: 
phonons and magnons. Here they are considered
in an approximate way as an additional quasi-static disorder: phonons - as positional
disorder, magnons - as spin-orientation  disorder~\cite{EMC+15}.  
In addition, we neglected the $T$-dependency of the Fermi-Dirac
statistics and identified the actual chemical potential $\mu(T)$  with $E_{\rm  F}$. The ability to treat
both thermal disorder sources within the CPA formalism made 
this approach especially convenient. 
Even though the non-local details and the specific features of the thermal oscillatory
modes are neglected, the practical use of this approach has been convincingly
demonstrated~\cite{EMC+15,CMME17,MPC+17}. 

The $T$-dependency of the mean amplitude of the atomic displacements 
was determined here by the Debye theory
(the effective Debye temperature was taken as an average over
atomic types), whereas the directions of displacements were selected
along the basis vectors to keep the conformity with
the lattice.  The atomic spins were assumed to have $T$-independent
amplitudes $m_i$ (${i=1,..,N}$; $N$ is the number of atoms in the
unit cell), and thus were calculated from first principles, but
the adequateness of the $T$-dependency of their angular distribution expressed by weights,
must be determined.
At a fixed temperature $T$, the angular distribution of the $i$-th atomic
spin gives its effective average value: 
${m_i\sum_{\nu}p_{i\nu}(T)\vec{e}_{\nu}=\langle\vec m_i\rangle(T)}$
($\{\vec{e}_{\nu}\}$  is a fixed set of all possible spatial
directions). The angular distribution was assumed to be
Gibbs-like (see Eqs.~13-15 in Ref.~\cite{EMC+15})
with weights $\{p_{i\nu}(T)\}$ determined by fitting the ex\-pe\-ri\-men\-tal
value: ${\langle\vec m_i\rangle(T)=\vec m_{i}^{\rm exp}(T)}$. Such a mapping is
unique only for a single magnetic sublattice, where the
experimental magnetization  unambiguously defines 
the angular distribution of each atomic spin, since ${\forall i: m^{\rm
    exp}_{i{\rm z}}=M_{\rm exp}/N}$ (index ``z'' denotes a projection on the
common magnetization axis). In the present case, even though $M_{\rm exp}(T)$ is known, the unit cell contains five
different magnetic sublattices: $i=$ V($4a$), Mn($4a$), V$(4b)$, Fe($8c$) and
Mn($8c$). We simplified this situation by assuming that the same form of
the $T$-dependency applies to all atomic spins that randomly share the same Wyckoff
site, which reduced the number of magnetic sublattices from five to three (i.e., $i=$ $4a$, $4b$ and $8c$). To avoid the remaining ambiguity, we assumed some reasonable form of the $T$-dependency for
each sublattice, e.g., by implying a sublattice-specific Bloch's law: 
${\langle m_{i\rm z}\rangle(T)=m_{i\rm z}\left(1-(T/T_{i})^{\alpha_i}\right)^{\beta_i}}$,
where $\alpha_i$, $\beta_i$  and $T_{i}$ (playing the role of an ordering
temperature for the $i$-th sublattice) are $T$-independent
fitting parameters and ${m_{i\rm z}=m_{i\rm z}(0)}$ - the ground-state
atomic spin moments calculated from first principles. Thus, we fitted
$M_{\rm exp}(T)$ by using the following expression:
${\sum_im_{i\rm z}\left(1-(T/T_{i})^{\alpha_i}\right)^{\beta_i}=M_{\rm fit}(T)\longrightarrow M_{\rm exp}(T)}$,~with $i$ running over 
three atomic sublattices entering the unit cell  with the corresponding
multiplicities, which are implicitly included in $m_{i\rm z}$. 
\begin{figure}
\centering
  \includegraphics[width=0.9\linewidth,clip]{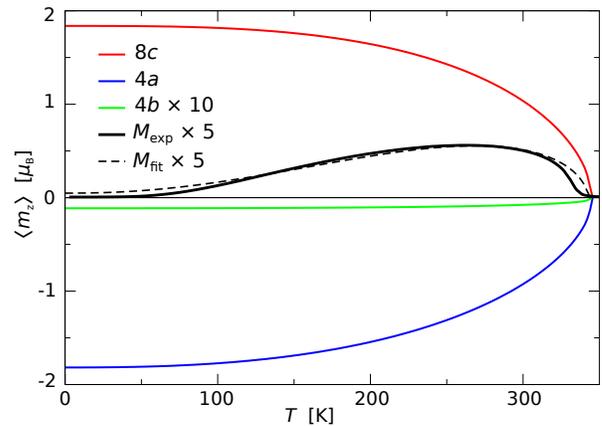}
 \caption{Experimental magnetization~\cite{SFC+17} in $\mu_{\rm B}$/f.u. (black solid line) versus fit
   (black dashed line) together with the z-projections of the sublattice spin
   moments derived from the fit (red, blue and green correspond to $8c$
   (Mn$_{0.5}$Fe$_{0.5}$), $4a$ (Mn$_{0.5}$V$_{0.5}$) and $4b$ (Al), respectively).
   \label{FIG:mag}} 
\end{figure}
The fit (see Fig.\,\ref{FIG:mag}) resulted in 
rather close sets for the ordering temperatures $T_i=$ 345.6, 345.0 and
345.0\,K, as well as for the power factors $\alpha_i=2.52$, 2.92 and
 3.10, $\beta_i=0.56$, 0.35 and 0.55 for sites $4a$, $4b$ and $8c$,
 respectively. These factors appeared to have the same order of magnitude as
 those in the conventional Bloch's law (${\alpha=\nicefrac{3}{2}}$,
 ${\beta=\nicefrac{1}{3}}$).

\begin{figure}
\centering
  \includegraphics[width=0.9\linewidth,clip]{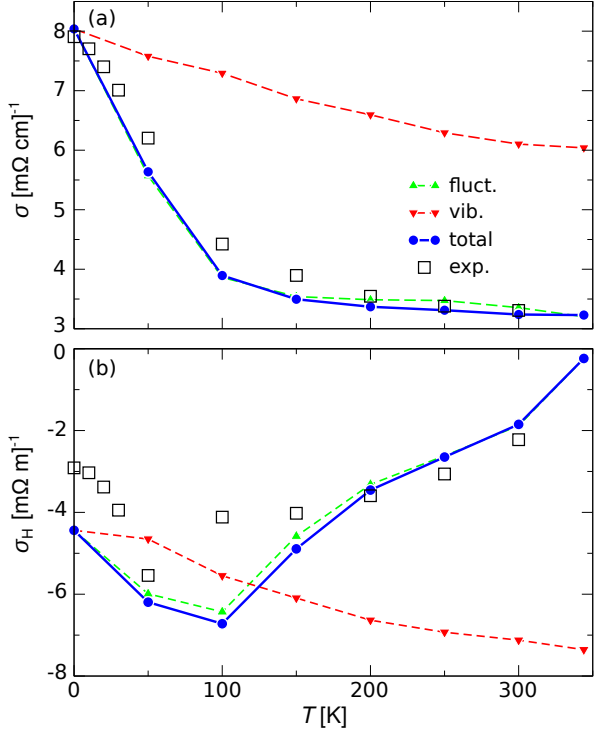}
 \caption{(a)~Longitudinal $\sigma$ and (b)~anomalous Hall
   $\sigma_{\rm H}$ conductivities. Dashed  green (up-triangles) and red (down-triangles) curves correspond to the case when
    only either magnetic fluctuation or atomic vibrations, respectively,
   are taken into account. The blue curve (circles) corresponds to the simultaneous
   inclusion of both scattering sources. Hollow squares stand for experimental values~\cite{SFC+17}.\label{FIG:transport}} 
\end{figure}

The conductivities $\sigma$ and $\sigma_{\rm H}$ calculated as functions of $T$ are shown in Fig.\,\ref{FIG:transport}.
The effects of spin-fluctuations and atomic vibrations are demonstrated by two additional curves, where
the calculation accounts either only for spin-fluctuations (marked as
``fluct.'') or only for atomic vibrations (``vib.''). These scattering sources cannot be combined,
 neither as parallel nor as sequential resistors (i.e., neither of these
 combinations gives the blue curve), even at low temperatures. On the
 other hand, the result based on the spin-fluctuations alone (green) followed the
total curve (blue) more closely indicating that the spin disorder is
the dominant scattering source. Both computed $\sigma$ and $\sigma_{\rm
  H}$  reasonably agreed with the 
experimental values over the whole temperature range. The strongest deviation from experiment 
 was simultaneously observed around 100\,K  for both quantities (in case
of $\sigma_{\rm H}$@100\,K the deviation was more than 50\,\%, however due to ${\sigma_{\rm H}/\sigma\sim 10^{-3}}$, for the absolute deviation the relation ${\delta\sigma_{\rm
    H}\sim\sigma_{\rm H}\ll\delta\sigma\ll\sigma}$ holds). The main reason for the
deviations are the aforementioned assumptions about the angular
distribution of the spin moments, which might deviate from the actual distribution more strongly in the $T$-range where the dispersion
is already large, but the distribution is still far
from uniform. The adequate description of this temperature regime
becomes rather complicated, but it can be improved by systematically considering different
aspects influencing the distribution of the local moments, such
as the specific features of the magnon dispersion, additional angular correlations
 imposed by relativistic effects, and possible longitudinal
 spin fluctuations. 

\begin{figure}
\centering
  \includegraphics[width=0.9\linewidth,clip]{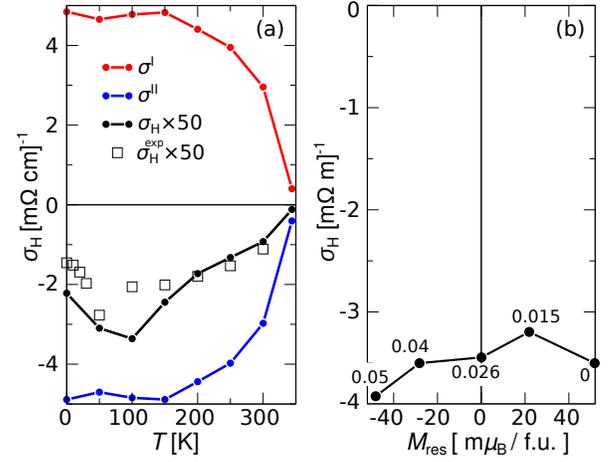}
 \caption{\label{FIG:final} (a)~Anomalous Hall conductivities versus
   temperature. Red and blue curves correspond to the Fermi-surface
   $\sigma^{\rm I}$ and Fermi-sea $\sigma^{\rm II}$ terms. The black
   solid line is their sum $\sigma_{\rm H}$, scaled up by a factor 
   50, as well as the experimental values~\cite{SFC+17} ($\sigma_{\rm H}^{\rm
     exp}$, hollow squares). (b)~$\sigma_{\rm H}$ vs residual magnetization
   $M_{\rm res}$. Annotated values correspond to the stoichiometric
   variation $\delta$ on $4a$ site, Mn$_{0.5+\delta}$V$_{0.5-\delta}$, which controls $M_{\rm res}$.} 
\end{figure}

In addition to the determination of the chemical order, the present calculations explain several aspects specific to 
ferrimagnets, such as the directions of local moments in the magnetically
compensated state. Since the reversal of local moment occurs
simultaneously with the sign change of $\sigma_{\rm H}$, ${\sigma_{\rm H}<0}$, the moments of Mn and Fe on $8c$ positions are
positive (aligned along an infinitesimal small external magnetic field),
whereas those of Mn and V on $4a$ are negative. 
Further, the residual chemical disorder is shown to reduce the  AHE: in the Kubo--Bastin formalism~\cite{KCE15}, the transverse conductivity is the sum of the
Fermi-surface term ($\sigma^{\rm I}$, the contribution from the conducting
electrons at $E_{\rm F}$) and the Fermi-sea term ($\sigma^{\rm II}$, the contribution from
the occupied states), ${\sigma_{\rm  H}=\sigma^{\rm I}+\sigma^{\rm
    II}\sim-10^{-5}\,(\mu\Omega\text{cm})^{-1}}$,  which appear to be
  large quantities with opposite sign: ${\sigma^{\rm  I}\sim-\sigma^{\rm
      II}\sim5\cdot10^{-3}\,(\mu\Omega\text{cm})^{-1}}$. This relation
  holds in the whole temperature range up to the magnetic critical point,
  where both terms simultaneously vanish (see Fig.\,\ref{FIG:final}\,a).   
While   $\sigma^{\rm II}$ is almost insensitive to the residual disorder,
  $\sigma^{\rm I}$ is strongly dependent on disorder and
 vanishes only close to the perfect limit, thereby increasing the total
 sum  $\sigma_{\rm H}$. However, this does not apply to the
  present material as it is strongly disordered. 

The non-trivial observation in which the ideally compensated collinear ferrimagnet can exhibit a
non-zero AHE does not unambiguously follow from the above data since
neither the experimental nor the theoretical situations are ideal. Small residual magnetization is present both in experiment and in the
ground-state calculations; the computed value is  ${M_{\rm res}=M_{\rm
  spin}+M_{\rm orbital}=0.0097+0.0421=0.05\,\mu_{\rm B}\text{/f.u.}}$ for
the nominal (non-shifted) $E_{\rm F}$. The applied calibration shift ${\Delta
  E_{\rm F}=20}$\,meV  slightly influences $M_{\rm res}$ further.  To demonstrate the nonzero AHE at ${M=0}$, we have adjusted the
stoichiometry so that ${M_{\rm res}=0}$ within the numerical
precision (see Fig.\,\ref{FIG:final}\,b). This can be achieved, for
instance, by a slight excess of Mn on the $4a$ site: Mn$_{0.5+\delta}$V$_{0.5-\delta}$. Some non-zero values of $M_{\rm res}$
are negative since we do not change the directions of the atomic
moments, in order to preserve the sign of $\sigma_{\rm H}$. As it
follows, $\sigma_{\rm H}$ continuously changes with
$M_{\rm res}$  and does not show any minimum in the
amplitude by approaching ${M_{\rm res}=0}$. Thus, the AHE should not vanish in the ferrimagnets
because of magnetic compensation. We emphasize that the aspect
of full compensation is rather fundamentally than technologically
relevant, since it is almost impossible in practice
to remove small rest of the magnetization even in antiferromagnets. On
the other hand, this makes the  verification of a non-vanishing AHE technically difficult,
since $\sigma_{\rm H}$ (or $\rho_{\rm H}$) changes its sign with the reversal of the external
magnetic field and thus passes through zero~\cite{NF65,SNF+17}. 

To conclude, we provided an extended first-principles description of the
temperature-dependent charge transport in the compensated ferrimagnet
Mn$_{1.5}$V$_{0.5}$FeAl, which was in good agreement with experiment. In
particular, we analyzed the influence of disorder on a charge transport and proved
the possibility of a  non-zero anomalous Hall effect in the ideally compensated state.


\begin{thebibliography}{30}
\expandafter\ifx\csname natexlab\endcsname\relax\def\natexlab#1{#1}\fi
\expandafter\ifx\csname bibnamefont\endcsname\relax
  \def\bibnamefont#1{#1}\fi
\expandafter\ifx\csname bibfnamefont\endcsname\relax
  \def\bibfnamefont#1{#1}\fi
\expandafter\ifx\csname citenamefont\endcsname\relax
  \def\citenamefont#1{#1}\fi
\expandafter\ifx\csname url\endcsname\relax
  \def\url#1{\texttt{#1}}\fi
\expandafter\ifx\csname urlprefix\endcsname\relax\def\urlprefix{URL }\fi
\providecommand{\bibinfo}[2]{#2}
\providecommand{\eprint}[2][]{\url{#2}}

\bibitem[{\citenamefont{MacDonald and Tsoi}(2011)}]{MT11}
\bibinfo{author}{\bibfnamefont{A.~H.} \bibnamefont{MacDonald}}
  \bibnamefont{and} \bibinfo{author}{\bibfnamefont{M.}~\bibnamefont{Tsoi}},
  \bibinfo{journal}{Phil. Trans. R. Soc. A} \textbf{\bibinfo{volume}{369}},
  \bibinfo{pages}{3098} (\bibinfo{year}{2011}).

\bibitem[{\citenamefont{Barthem et~al.}(2013)\citenamefont{Barthem, Colin,
  Mayaffre, Julien, and Givord}}]{BCM+13}
\bibinfo{author}{\bibfnamefont{V.~M. T.~S.} \bibnamefont{Barthem}},
  \bibinfo{author}{\bibfnamefont{C.~V.} \bibnamefont{Colin}},
  \bibinfo{author}{\bibfnamefont{H.}~\bibnamefont{Mayaffre}},
  \bibinfo{author}{\bibfnamefont{M.-H.} \bibnamefont{Julien}},
  \bibnamefont{and} \bibinfo{author}{\bibfnamefont{D.}~\bibnamefont{Givord}},
  \bibinfo{journal}{Nat. Commun.} \textbf{\bibinfo{volume}{4}},
  \bibinfo{pages}{2892} (\bibinfo{year}{2013}).

\bibitem[{\citenamefont{Wadley et~al.}(2016)\citenamefont{Wadley, Howells, {\v
  Z}elezn{\'y}, Andrews, Hills, Campion, Nov{\'a}k, Olejn{\'\i}k, Maccherozzi,
  Dhesi et~al.}}]{WHZ+16}
\bibinfo{author}{\bibfnamefont{P.}~\bibnamefont{Wadley}},
  \bibinfo{author}{\bibfnamefont{B.}~\bibnamefont{Howells}},
  \bibinfo{author}{\bibfnamefont{J.}~\bibnamefont{{\v Z}elezn{\'y}}},
  \bibinfo{author}{\bibfnamefont{C.}~\bibnamefont{Andrews}},
  \bibinfo{author}{\bibfnamefont{V.}~\bibnamefont{Hills}},
  \bibinfo{author}{\bibfnamefont{R.~P.} \bibnamefont{Campion}},
  \bibinfo{author}{\bibfnamefont{V.}~\bibnamefont{Nov{\'a}k}},
  \bibinfo{author}{\bibfnamefont{K.}~\bibnamefont{Olejn{\'\i}k}},
  \bibinfo{author}{\bibfnamefont{F.}~\bibnamefont{Maccherozzi}},
  \bibinfo{author}{\bibfnamefont{S.~S.} \bibnamefont{Dhesi}},
  \bibnamefont{et~al.}, \bibinfo{journal}{Science}
  \textbf{\bibinfo{volume}{351}}, \bibinfo{pages}{192404}
  (\bibinfo{year}{2016}).

\bibitem[{\citenamefont{van~der Bijl et~al.}(2013)\citenamefont{van~der Bijl,
  Troncoso, and Duine}}]{BTD13}
\bibinfo{author}{\bibfnamefont{E.}~\bibnamefont{van~der Bijl}},
  \bibinfo{author}{\bibfnamefont{R.~E.} \bibnamefont{Troncoso}},
  \bibnamefont{and} \bibinfo{author}{\bibfnamefont{R.~A.} \bibnamefont{Duine}},
  \bibinfo{journal}{Phys. Rev. B} \textbf{\bibinfo{volume}{88}},
  \bibinfo{pages}{064417} (\bibinfo{year}{2013}).

\bibitem[{\citenamefont{Watanabe et~al.}(2016)\citenamefont{Watanabe, Hoshi,
  and Ohe}}]{WHO16}
\bibinfo{author}{\bibfnamefont{H.}~\bibnamefont{Watanabe}},
  \bibinfo{author}{\bibfnamefont{K.}~\bibnamefont{Hoshi}}, \bibnamefont{and}
  \bibinfo{author}{\bibfnamefont{J.-I.} \bibnamefont{Ohe}},
  \bibinfo{journal}{Phys. Rev. B} \textbf{\bibinfo{volume}{94}},
  \bibinfo{pages}{125143} (\bibinfo{year}{2016}).

\bibitem[{\citenamefont{Singh et~al.}(2016)\citenamefont{Singh, D'Souza, Nayak,
  Suard, Chapon, Senyshyn, Petricek, Skourski, Nicklas, Felser
  et~al.}}]{SSN+16}
\bibinfo{author}{\bibfnamefont{S.}~\bibnamefont{Singh}},
  \bibinfo{author}{\bibfnamefont{S.~W.} \bibnamefont{D'Souza}},
  \bibinfo{author}{\bibfnamefont{J.}~\bibnamefont{Nayak}},
  \bibinfo{author}{\bibfnamefont{E.}~\bibnamefont{Suard}},
  \bibinfo{author}{\bibfnamefont{L.}~\bibnamefont{Chapon}},
  \bibinfo{author}{\bibfnamefont{A.}~\bibnamefont{Senyshyn}},
  \bibinfo{author}{\bibfnamefont{V.}~\bibnamefont{Petricek}},
  \bibinfo{author}{\bibfnamefont{Y.}~\bibnamefont{Skourski}},
  \bibinfo{author}{\bibfnamefont{M.}~\bibnamefont{Nicklas}},
  \bibinfo{author}{\bibfnamefont{C.}~\bibnamefont{Felser}},
  \bibnamefont{et~al.}, \bibinfo{journal}{Nat. Commun.}
  \textbf{\bibinfo{volume}{7}}, \bibinfo{pages}{12671} (\bibinfo{year}{2016}).

\bibitem[{\citenamefont{K\"{u}bler and Felser}(2014)}]{KF14}
\bibinfo{author}{\bibfnamefont{J.}~\bibnamefont{K\"{u}bler}} \bibnamefont{and}
  \bibinfo{author}{\bibfnamefont{C.}~\bibnamefont{Felser}},
  \bibinfo{journal}{Europhys. Lett.} \textbf{\bibinfo{volume}{108}},
  \bibinfo{pages}{67001} (\bibinfo{year}{2014}).

\bibitem[{\citenamefont{Feng et~al.}(2015)\citenamefont{Feng, Guo, Zhou, Yao,
  and Niu}}]{FGZ+15}
\bibinfo{author}{\bibfnamefont{W.}~\bibnamefont{Feng}},
  \bibinfo{author}{\bibfnamefont{G.-Y.} \bibnamefont{Guo}},
  \bibinfo{author}{\bibfnamefont{J.}~\bibnamefont{Zhou}},
  \bibinfo{author}{\bibfnamefont{Y.}~\bibnamefont{Yao}}, \bibnamefont{and}
  \bibinfo{author}{\bibfnamefont{Q.}~\bibnamefont{Niu}},
  \bibinfo{journal}{Phys. Rev. B} \textbf{\bibinfo{volume}{92}},
  \bibinfo{pages}{144426} (\bibinfo{year}{2015}).

\bibitem[{\citenamefont{Chadov et~al.}(2013)\citenamefont{Chadov, Kiss, and
  Felser}}]{CKF13}
\bibinfo{author}{\bibfnamefont{S.}~\bibnamefont{Chadov}},
  \bibinfo{author}{\bibfnamefont{J.}~\bibnamefont{Kiss}}, \bibnamefont{and}
  \bibinfo{author}{\bibfnamefont{C.}~\bibnamefont{Felser}},
  \bibinfo{journal}{Adv.~Func.~Mater.} \textbf{\bibinfo{volume}{23}},
  \bibinfo{pages}{832} (\bibinfo{year}{2013}).

\bibitem[{\citenamefont{Wollmann
  et~al.}(2015{\natexlab{a}})\citenamefont{Wollmann, Fecher, Chadov, and
  Felser}}]{WFCF15}
\bibinfo{author}{\bibfnamefont{L.}~\bibnamefont{Wollmann}},
  \bibinfo{author}{\bibfnamefont{G.~H.} \bibnamefont{Fecher}},
  \bibinfo{author}{\bibfnamefont{S.}~\bibnamefont{Chadov}}, \bibnamefont{and}
  \bibinfo{author}{\bibfnamefont{C.}~\bibnamefont{Felser}},
  \bibinfo{journal}{J. Phys. D: Appl. Phys.} \textbf{\bibinfo{volume}{48}},
  \bibinfo{pages}{164004} (\bibinfo{year}{2015}{\natexlab{a}}).

\bibitem[{\citenamefont{Chadov et~al.}(2015)\citenamefont{Chadov, D'Souza,
  Wollmann, Kiss, Fecher, and Felser}}]{CSW+15}
\bibinfo{author}{\bibfnamefont{S.}~\bibnamefont{Chadov}},
  \bibinfo{author}{\bibfnamefont{S.~W.} \bibnamefont{D'Souza}},
  \bibinfo{author}{\bibfnamefont{L.}~\bibnamefont{Wollmann}},
  \bibinfo{author}{\bibfnamefont{J.}~\bibnamefont{Kiss}},
  \bibinfo{author}{\bibfnamefont{G.~H.} \bibnamefont{Fecher}},
  \bibnamefont{and} \bibinfo{author}{\bibfnamefont{C.}~\bibnamefont{Felser}},
  \bibinfo{journal}{Phys. Rev. B} \textbf{\bibinfo{volume}{91}},
  \bibinfo{pages}{094203} (\bibinfo{year}{2015}).

\bibitem[{\citenamefont{Chen et~al.}(2014)\citenamefont{Chen, Niu, and
  MacDonald}}]{CNM14}
\bibinfo{author}{\bibfnamefont{H.}~\bibnamefont{Chen}},
  \bibinfo{author}{\bibfnamefont{Q.}~\bibnamefont{Niu}}, \bibnamefont{and}
  \bibinfo{author}{\bibfnamefont{A.~H.} \bibnamefont{MacDonald}},
  \bibinfo{journal}{Phys. Rev. Lett.} \textbf{\bibinfo{volume}{112}},
  \bibinfo{pages}{017205} (\bibinfo{year}{2014}).

\bibitem[{\citenamefont{Kleiner}(1966)}]{Kle66}
\bibinfo{author}{\bibfnamefont{W.~H.} \bibnamefont{Kleiner}},
  \bibinfo{journal}{Phys. Rev.} \textbf{\bibinfo{volume}{142}},
  \bibinfo{pages}{318} (\bibinfo{year}{1966}).

\bibitem[{\citenamefont{Seemann et~al.}(2015)\citenamefont{Seemann,
  K\"odderitzsch, Wimmer, and Ebert}}]{SKWE15}
\bibinfo{author}{\bibfnamefont{M.}~\bibnamefont{Seemann}},
  \bibinfo{author}{\bibfnamefont{D.}~\bibnamefont{K\"odderitzsch}},
  \bibinfo{author}{\bibfnamefont{S.}~\bibnamefont{Wimmer}}, \bibnamefont{and}
  \bibinfo{author}{\bibfnamefont{H.}~\bibnamefont{Ebert}},
  \bibinfo{journal}{Phys. Rev. B} \textbf{\bibinfo{volume}{92}},
  \bibinfo{pages}{155138} (\bibinfo{year}{2015}).

\bibitem[{\citenamefont{Slater}(1936)}]{Sla36}
\bibinfo{author}{\bibfnamefont{J.~C.} \bibnamefont{Slater}},
  \bibinfo{journal}{Phys. Rev.} \textbf{\bibinfo{volume}{49}},
  \bibinfo{pages}{931} (\bibinfo{year}{1936}).

\bibitem[{\citenamefont{Pauling}(1938)}]{Pau38}
\bibinfo{author}{\bibfnamefont{L.}~\bibnamefont{Pauling}},
  \bibinfo{journal}{Phys. Rev.} \textbf{\bibinfo{volume}{54}},
  \bibinfo{pages}{899} (\bibinfo{year}{1938}).

\bibitem[{\citenamefont{Li et~al.}(2013)\citenamefont{Li, Liu, Zhang, Du,
  Zhang, Wang, and Wu}}]{LLZ+12}
\bibinfo{author}{\bibfnamefont{G.~J.} \bibnamefont{Li}},
  \bibinfo{author}{\bibfnamefont{E.~K.} \bibnamefont{Liu}},
  \bibinfo{author}{\bibfnamefont{Y.~J.} \bibnamefont{Zhang}},
  \bibinfo{author}{\bibfnamefont{Y.}~\bibnamefont{Du}},
  \bibinfo{author}{\bibfnamefont{H.~W.} \bibnamefont{Zhang}},
  \bibinfo{author}{\bibfnamefont{W.~H.} \bibnamefont{Wang}}, \bibnamefont{and}
  \bibinfo{author}{\bibfnamefont{G.~H.} \bibnamefont{Wu}}, \bibinfo{journal}{J.
  Appl. Phys.} \textbf{\bibinfo{volume}{113}}, \bibinfo{pages}{103903}
  (\bibinfo{year}{2013}).

\bibitem[{\citenamefont{Wollmann
  et~al.}(2015{\natexlab{b}})\citenamefont{Wollmann, Chadov, K\"ubler, and
  Felser}}]{WCKF15}
\bibinfo{author}{\bibfnamefont{L.}~\bibnamefont{Wollmann}},
  \bibinfo{author}{\bibfnamefont{S.}~\bibnamefont{Chadov}},
  \bibinfo{author}{\bibfnamefont{J.}~\bibnamefont{K\"ubler}}, \bibnamefont{and}
  \bibinfo{author}{\bibfnamefont{C.}~\bibnamefont{Felser}},
  \bibinfo{journal}{Phys. Rev. B} \textbf{\bibinfo{volume}{92}},
  \bibinfo{pages}{064417} (\bibinfo{year}{2015}{\natexlab{b}}).

\bibitem[{\citenamefont{Nayak et~al.}(2015)\citenamefont{Nayak, Nicklas,
  Chadov, Khuntia, Shekhar, Kalache, Baenitz, Skourski, Guduru, Puri
  et~al.}}]{NNC+15}
\bibinfo{author}{\bibfnamefont{A.~K.} \bibnamefont{Nayak}},
  \bibinfo{author}{\bibfnamefont{M.}~\bibnamefont{Nicklas}},
  \bibinfo{author}{\bibfnamefont{S.}~\bibnamefont{Chadov}},
  \bibinfo{author}{\bibfnamefont{P.}~\bibnamefont{Khuntia}},
  \bibinfo{author}{\bibfnamefont{C.}~\bibnamefont{Shekhar}},
  \bibinfo{author}{\bibfnamefont{A.}~\bibnamefont{Kalache}},
  \bibinfo{author}{\bibfnamefont{M.}~\bibnamefont{Baenitz}},
  \bibinfo{author}{\bibfnamefont{Y.}~\bibnamefont{Skourski}},
  \bibinfo{author}{\bibfnamefont{V.~K.} \bibnamefont{Guduru}},
  \bibinfo{author}{\bibfnamefont{A.}~\bibnamefont{Puri}}, \bibnamefont{et~al.},
  \bibinfo{journal}{Nat. Mater.} \textbf{\bibinfo{volume}{14}},
  \bibinfo{pages}{679} (\bibinfo{year}{2015}).

\bibitem[{\citenamefont{Stinshoff
  et~al.}(2017{\natexlab{a}})\citenamefont{Stinshoff, Nayak, Fecher, Balke,
  Ouardi, Skourski, Nakamura, and Felser}}]{SNF+17}
\bibinfo{author}{\bibfnamefont{R.}~\bibnamefont{Stinshoff}},
  \bibinfo{author}{\bibfnamefont{A.~K.} \bibnamefont{Nayak}},
  \bibinfo{author}{\bibfnamefont{G.~H.} \bibnamefont{Fecher}},
  \bibinfo{author}{\bibfnamefont{B.}~\bibnamefont{Balke}},
  \bibinfo{author}{\bibfnamefont{S.}~\bibnamefont{Ouardi}},
  \bibinfo{author}{\bibfnamefont{Y.}~\bibnamefont{Skourski}},
  \bibinfo{author}{\bibfnamefont{T.}~\bibnamefont{Nakamura}}, \bibnamefont{and}
  \bibinfo{author}{\bibfnamefont{C.}~\bibnamefont{Felser}},
  \bibinfo{journal}{Phys. Rev. B} \textbf{\bibinfo{volume}{95}},
  \bibinfo{pages}{060410} (\bibinfo{year}{2017}{\natexlab{a}}).

\bibitem[{\citenamefont{Stinshoff
  et~al.}(2017{\natexlab{b}})\citenamefont{Stinshoff, Fecher, Chadov, Nayak,
  Balke, Ouardi, Nakamura, and Felser}}]{SFC+17}
\bibinfo{author}{\bibfnamefont{R.}~\bibnamefont{Stinshoff}},
  \bibinfo{author}{\bibfnamefont{G.~H.} \bibnamefont{Fecher}},
  \bibinfo{author}{\bibfnamefont{S.}~\bibnamefont{Chadov}},
  \bibinfo{author}{\bibfnamefont{A.~K.} \bibnamefont{Nayak}},
  \bibinfo{author}{\bibfnamefont{B.}~\bibnamefont{Balke}},
  \bibinfo{author}{\bibfnamefont{S.}~\bibnamefont{Ouardi}},
  \bibinfo{author}{\bibfnamefont{T.}~\bibnamefont{Nakamura}}, \bibnamefont{and}
  \bibinfo{author}{\bibfnamefont{C.}~\bibnamefont{Felser}},
  \bibinfo{journal}{accepted in AIP~Advances}
  (\bibinfo{year}{2017}{\natexlab{b}}).

\bibitem[{\citenamefont{Perdew et~al.}(1996)\citenamefont{Perdew, Burke, and
  Ernzerhof}}]{PBE96}
\bibinfo{author}{\bibfnamefont{J.~P.} \bibnamefont{Perdew}},
  \bibinfo{author}{\bibfnamefont{K.}~\bibnamefont{Burke}}, \bibnamefont{and}
  \bibinfo{author}{\bibfnamefont{M.}~\bibnamefont{Ernzerhof}},
  \bibinfo{journal}{Phys. Rev. Lett.} \textbf{\bibinfo{volume}{77}},
  \bibinfo{pages}{3865} (\bibinfo{year}{1996}).

\bibitem[{\citenamefont{Ebert et~al.}(2011)\citenamefont{Ebert, K\"odderitzsch,
  and Min\'{a}r}}]{EKM11}
\bibinfo{author}{\bibfnamefont{H.}~\bibnamefont{Ebert}},
  \bibinfo{author}{\bibfnamefont{D.}~\bibnamefont{K\"odderitzsch}},
  \bibnamefont{and}
  \bibinfo{author}{\bibfnamefont{J.}~\bibnamefont{Min\'{a}r}},
  \bibinfo{journal}{Rep. Prog. Phys.} \textbf{\bibinfo{volume}{74}},
  \bibinfo{pages}{096501} (\bibinfo{year}{2011}).

\bibitem[{\citenamefont{K\"{o}dderitzsch
  et~al.}(2015)\citenamefont{K\"{o}dderitzsch, Chadova, and Ebert}}]{KCE15}
\bibinfo{author}{\bibfnamefont{D.}~\bibnamefont{K\"{o}dderitzsch}},
  \bibinfo{author}{\bibfnamefont{K.}~\bibnamefont{Chadova}}, \bibnamefont{and}
  \bibinfo{author}{\bibfnamefont{H.}~\bibnamefont{Ebert}},
  \bibinfo{journal}{Phys. Rev. B} \textbf{\bibinfo{volume}{92}},
  \bibinfo{pages}{184415} (\bibinfo{year}{2015}).

\bibitem[{\citenamefont{Soven}(1967)}]{Sov67}
\bibinfo{author}{\bibfnamefont{P.}~\bibnamefont{Soven}},
  \bibinfo{journal}{Phys. Rev.} \textbf{\bibinfo{volume}{156}},
  \bibinfo{pages}{809} (\bibinfo{year}{1967}).

\bibitem[{\citenamefont{Taylor}(1967)}]{Tay67}
\bibinfo{author}{\bibfnamefont{D.~W.} \bibnamefont{Taylor}},
  \bibinfo{journal}{Phys. Rev.} \textbf{\bibinfo{volume}{156}},
  \bibinfo{pages}{1017} (\bibinfo{year}{1967}).

\bibitem[{\citenamefont{Ebert et~al.}(2015)\citenamefont{Ebert, Mankovsky,
  Chadova, Polesya, Min\'ar, and K\"odderitzsch}}]{EMC+15}
\bibinfo{author}{\bibfnamefont{H.}~\bibnamefont{Ebert}},
  \bibinfo{author}{\bibfnamefont{S.}~\bibnamefont{Mankovsky}},
  \bibinfo{author}{\bibfnamefont{K.}~\bibnamefont{Chadova}},
  \bibinfo{author}{\bibfnamefont{S.}~\bibnamefont{Polesya}},
  \bibinfo{author}{\bibfnamefont{J.}~\bibnamefont{Min\'ar}}, \bibnamefont{and}
  \bibinfo{author}{\bibfnamefont{D.}~\bibnamefont{K\"odderitzsch}},
  \bibinfo{journal}{Phys. Rev. B} \textbf{\bibinfo{volume}{91}},
  \bibinfo{pages}{165132} (\bibinfo{year}{2015}).

\bibitem[{\citenamefont{Chadova et~al.}(2017)\citenamefont{Chadova, Mankovsky,
  Min\'ar, and Ebert}}]{CMME17}
\bibinfo{author}{\bibfnamefont{K.}~\bibnamefont{Chadova}},
  \bibinfo{author}{\bibfnamefont{S.}~\bibnamefont{Mankovsky}},
  \bibinfo{author}{\bibfnamefont{J.}~\bibnamefont{Min\'ar}}, \bibnamefont{and}
  \bibinfo{author}{\bibfnamefont{H.}~\bibnamefont{Ebert}},
  \bibinfo{journal}{Phys. Rev. B} \textbf{\bibinfo{volume}{95}},
  \bibinfo{pages}{125109} (\bibinfo{year}{2017}).

\bibitem[{\citenamefont{Mankovsky et~al.}(2017)\citenamefont{Mankovsky,
  Polesya, Chadova, Ebert, Staunton, Gruenbaum, Schoen, Back, Chen, and
  Song}}]{MPC+17}
\bibinfo{author}{\bibfnamefont{S.}~\bibnamefont{Mankovsky}},
  \bibinfo{author}{\bibfnamefont{S.}~\bibnamefont{Polesya}},
  \bibinfo{author}{\bibfnamefont{K.}~\bibnamefont{Chadova}},
  \bibinfo{author}{\bibfnamefont{H.}~\bibnamefont{Ebert}},
  \bibinfo{author}{\bibfnamefont{J.~B.} \bibnamefont{Staunton}},
  \bibinfo{author}{\bibfnamefont{T.}~\bibnamefont{Gruenbaum}},
  \bibinfo{author}{\bibfnamefont{M.~A.~W.} \bibnamefont{Schoen}},
  \bibinfo{author}{\bibfnamefont{C.~H.} \bibnamefont{Back}},
  \bibinfo{author}{\bibfnamefont{X.~Z.} \bibnamefont{Chen}}, \bibnamefont{and}
  \bibinfo{author}{\bibfnamefont{C.}~\bibnamefont{Song}},
  \bibinfo{journal}{Phys. Rev. B} \textbf{\bibinfo{volume}{95}},
  \bibinfo{pages}{155139} (\bibinfo{year}{2017}).

\bibitem[{\citenamefont{Novogrudskii and Fakidov}(1965)}]{NF65}
\bibinfo{author}{\bibfnamefont{V.~N.} \bibnamefont{Novogrudskii}}
  \bibnamefont{and} \bibinfo{author}{\bibfnamefont{I.~G.}
  \bibnamefont{Fakidov}}, \bibinfo{journal}{Sov. Phys. J.E.T.P.}
  \textbf{\bibinfo{volume}{47}}, \bibinfo{pages}{20} (\bibinfo{year}{1965}).

\end{thebibliography}

\end{document}